\documentclass[
reprint,
superscriptaddress,
showpacs,
showkeys,
amsmath, amssymb,
aps,
prl,
]{revtex4-1}

\usepackage{xcolor}
\usepackage{graphicx}
\usepackage{dcolumn}
\usepackage{bm}
\usepackage[colorlinks=true, allcolors=blue]{hyperref}

\graphicspath{{figures/}}
\usepackage[
separate-uncertainty=false,
range-units = single,
]{siunitx}

\usepackage{xspace}
\newcommand{\na}{$^{23}$Na\xspace}
\newcommand{\he}{$^{4}$He\xspace}

\newcommand{\x}{$x$\xspace}
\newcommand{\y}{$y$\xspace}
\newcommand{\z}{$z$\xspace}

\newcommand{\ket}[1]{\vert#1\rangle}

\usepackage{xstring}
\newcommand*{\aref}[1]{%
	\IfBeginWith{#1}{eq:}{Eq.~\ref{#1}}{}
	\IfBeginWith{#1}{fig:}{Fig.~\ref{#1}}{}%
	\IfBeginWith{#1}{tab:}{Table~\ref{#1}}{}%
	\IfBeginWith{#1}{appendix:}{Appendix~\ref{#1}}{}%
	\IfBeginWith{#1}{sec:}{Section~\ref{#1}}{}%
}

\begin{document}

\title{Measurement of the Canonical Equation of State of a Weakly Interacting 3D Bose Gas}
\date{\today}

\author{C. Mordini}
\thanks{These two authors contributed equally.}
\affiliation{INO-CNR BEC Center and Dipartimento di Fisica, Universit\`a di Trento, 38123 Povo, Italy}
\author{D. Trypogeorgos}
\thanks{These two authors contributed equally.}
\affiliation{INO-CNR BEC Center and Dipartimento di Fisica, Universit\`a di Trento, 38123 Povo, Italy}
\affiliation{Trento Institute for Fundamental Physics and Applications, INFN, 38123 Povo, Italy}
\author{A. Farolfi}
\affiliation{INO-CNR BEC Center and Dipartimento di Fisica, Universit\`a di Trento, 38123 Povo, Italy}
\affiliation{Trento Institute for Fundamental Physics and Applications, INFN, 38123 Povo, Italy}
\author{L. Wolswijk}
\affiliation{INO-CNR BEC Center and Dipartimento di Fisica, Universit\`a di Trento, 38123 Povo, Italy}
\affiliation{Trento Institute for Fundamental Physics and Applications, INFN, 38123 Povo, Italy}
\author{S. Stringari}
\affiliation{INO-CNR BEC Center and Dipartimento di Fisica, Universit\`a di Trento, 38123 Povo, Italy}
\author{G. Lamporesi}
\author{G. Ferrari}
\affiliation{INO-CNR BEC Center and Dipartimento di Fisica, Universit\`a di Trento, 38123 Povo, Italy}
\affiliation{Trento Institute for Fundamental Physics and Applications, INFN, 38123 Povo, Italy}
\email[]{gabriele.ferrari@unitn.it}
\homepage[]{http://bec.science.unitn.it}

\begin{abstract}
Using a multiple-image reconstruction method applied to a harmonically trapped Bose gas, we determine the equation of state of uniform matter across the critical transition point, within the local density approximation. Our experimental results provide the canonical description of pressure as a function of the specific volume, emphasizing the dramatic deviations from the ideal Bose gas behavior caused by interactions. They also provide clear evidence for the non-monotonic behavior with temperature of the chemical potential, which is a consequence of superfluidity.
The measured thermodynamic quantities are compared to mean-field predictions available for the interacting Bose gas. The limits of applicability of the local density approximation near the critical point are also discussed, focusing on the behavior of the isothermal compressibility.
\end{abstract}


\maketitle

{\it Introduction}. Although 25 years have passed since the first realization of a Bose--Einstein condensate (BEC) in a dilute gas of alkali atoms, the experimental investigation of the equation of state (EoS) of a weakly interacting Bose gas is still rather incomplete. The EoS of the ideal Bose gas (IBG) predicts peculiar features at finite temperature, e.g., saturation of the thermal component and infinite compressibility in the BEC phase, so it is of major importance to have a direct experimental access to the crucial role of interactions which violate the IBG behavior. Experiments at finite temperature have focused on the role of interactions on the temperature dependence of the BEC fraction \cite{Smith2011,Tammuz2011} and on the value of the critical temperature in both harmonically trapped and uniform configurations \cite{Gerbier2004,Gotlibovych2014}. Results on the EoS of both 3D \cite{Nascimbene2010,Meppelink2010} and 2D \cite{Yefsah2011,Desbuquois2014} Bose gases have been obtained in the framework of the grand canonical approach, where the pressure of the uniform gas is expressed in terms of the chemical potential. At zero temperature the above approach has proven successful in identifying the Lee--Huang--Yang correction to the EoS originating from beyond-mean-field quantum fluctuations \cite{Navon2011}.

Atomic samples trapped by non-uniform potentials can be used to extract the thermodynamic behavior of uniform matter through the use of the local density approximation (LDA) \cite{Cheng2007,Ho2010}. In 3D, the pressure is extracted from the measured column density of the trapped gas using the Gibbs--Duhem relation, while the chemical potential is usually obtained fitting the density distribution of the sample, with the exception of the unitary Fermi gas were the model-dependent measurement of the chemical potential was successfully avoided by measuring the compressibility of the gas \cite{Ku563}.

In this Letter, we obtain the EoS of a uniform, 3D, weakly interacting Bose gas at constant temperature $T$ using the LDA method. We measure the density profile of a trapped atomic sample exploiting the axial symmetry of the trapping potential through the inverse Abel transform \cite{Shin2006}. This provides direct access to the canonical formulation of the EoS.

The canonical and grand canonical descriptions are in principle equivalent in the thermodynamic limit: the density of the system, which is the key variable of the canonical picture, can be derived starting from measurements of grand canonical variables with the use of fundamental thermodynamic relations. Experimentally, however, this procedure is technically demanding in 3D Bose gases and has never been realized so far. Here we circumvent this through a direct measurement of the density of the trapped gas. This allows us to explore important features of the system evident in the canonical formulation, like the behavior of the pressure $p(v,T)$ at fixed temperature $T$ as a function of the specific volume $v=1/n$, and the non-monotonic behavior of the chemical potential $\mu$ as a function of the reduced temperature $T/T_c$, which is a direct consequence of superfluidity \cite{Papoular2012}. We note that the thermodynamics of the 3D Bose gas is not universal, but it depends on the specific strength of atomic interactions fixed by the scattering length $a$. Here we investigate the behavior of the EoS at constant $a$. Fixing $a$ and $T$, we explore the thermodynamics as a function of the density $n$.

The density of a 3D condensed gas spans several orders of magnitude from the visible thermal tails to the dense condensate center, requiring an imaging method with a much higher dynamic range than usual absorption imaging. We tackle this using partial-transfer absorption imaging (PTAI)~\cite{Freilich2010,Ramanathan2012} and a reconstruction method that produces highly accurate spatial profiles even for very dense samples \cite{Mordini2020PTAI}.

{\it Experimental procedure.} We produce partially condensed \na gases confined in a Ioffe--Pritchard trap with axial (radial) trapping frequency $\omega_x / 2\pi = \SI{8.83 \pm 0.02}{\hertz}$ ($\omega_\rho / 2\pi = \SI{100.8 \pm 0.7}{\hertz}$), where we let the BEC equilibrate for \SI{2}{\second} after the end of the evaporation ramp. We then extract a few percent of the atoms and image them {\it in-situ} along the vertical direction \z, obtaining an image of a tunable fraction of the column density $n_1(x,y) = \int n \,dz$.
We implement PTAI by radiating the sample with microwaves of Rabi frequency $\Omega / 2\pi = \SI{60.7 \pm 0.2}{\kilo\hertz}$ to outcouple a fraction of the atoms from $\ket{F, m_F} = \ket{1,-1}$, where they are magnetically trapped, to $\ket{2, -2}$ in the upper hyperfine manifold. The extracted atoms are subsequently imaged with $\pi$-polarized light resonant with the $F=2 \to F'=3$ cycling transition using a 5-\si{\micro\second}-long probe pulse with $I/I_{sat} = 4$, where $I_{sat}$ is the saturation intensity of \na~\cite{Horikoshi2017}. This process takes only a few microseconds and so it does not suffer from any losses due to spin flipping collisions, which are further suppressed by conservation of angular momentum.
We then release the remaining atoms from the trap and image them along \y after a time of flight of \SI{50}{\milli\second}. Imaging at high intensity allows us to calibrate the absorption cross section, obtaining an absolute measure of the atomic density~\cite{Reinaudi2007}. Details on the imaging calibration are provided in the SM.

\begin{figure}[t]
 \centering
 \includegraphics{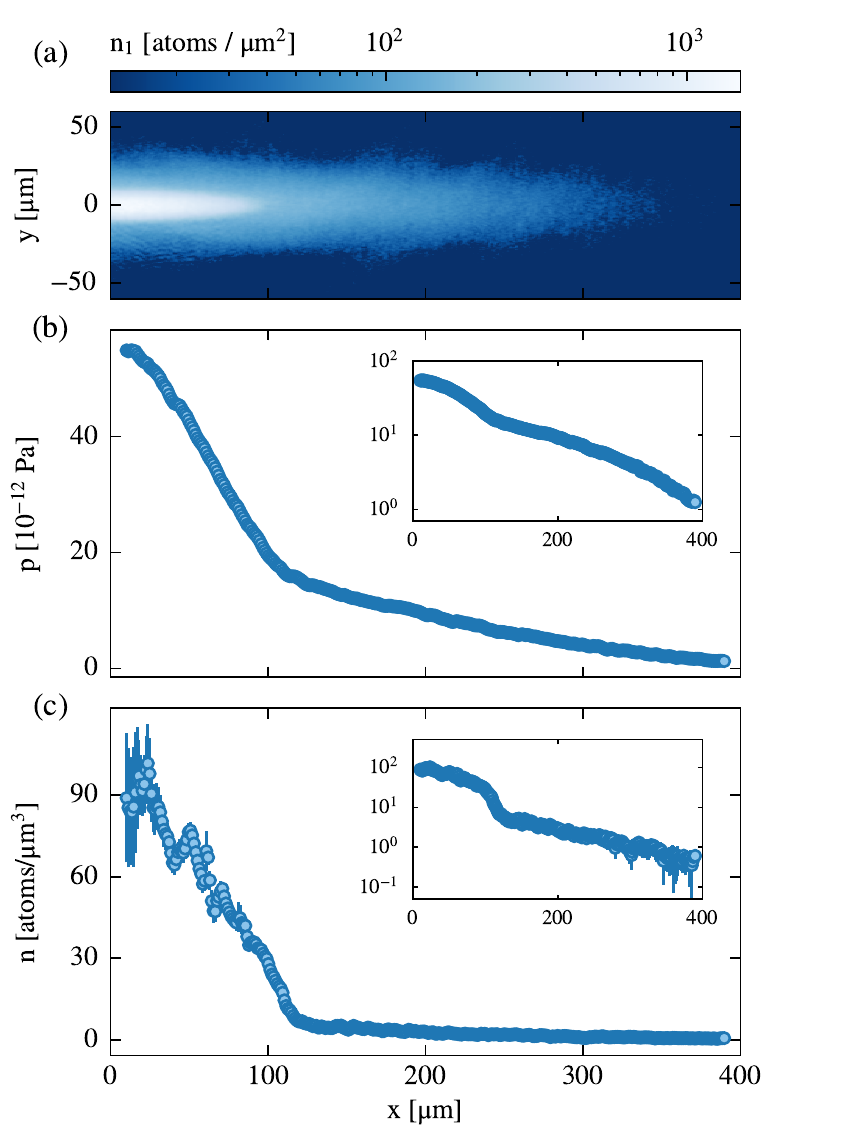}
 \caption{
 (a) Column density of the trapped sample, reconstructed from different partial extractions. Each pixel in the image results from the average of \SIrange{5}{80}{} images at different extraction ratios.
 (b) Pressure along the \x axis of the sample, obtained integrating $n_1$ along \y. The errorbars on the pressure are smaller than the marker size.
 (c) Axial density profile of the sample $n(x)$ obtained via the inverse Abel transform.
 The insets in (b) and (c) show pressure and density of the gas in log-scale, highlighting the high dynamic range needed to capture the different regimes of the Bose gas thermodynamics.
 }
 \label{fig:fig1}
\end{figure}

The reconstruction method combines multiple partial-transfer images to obtain a high-dynamic-range image of the column density. \textit{In-situ} measurements of $n_1$ are done for several microwave pulse times $\tau$ extracting a fraction $\sin^2(\Omega \tau / 2)$ of the atomic sample each time.
Long pulses, between \SIrange{1.5}{2.5}{\us} (extracting \SIrange{8}{20}{\percent}), yield a saturated image of the condensed part but allow to image the thermal tails with high signal-to-noise ratio. Short pulses, of \SIrange{0.5}{1.5}{\us} (\SIrange{1}{8}{\percent}), lead to an accurate image of the denser BEC core but the thermal tails are no longer visible.
In both cases the spectrum of the microwave pulse is broad enough to neglect the spatial detuning due to the trapping magnetic field and to consider extractions as uniform. We crop each image at a threshold set by the imaging conditions to retain only the non-saturated region, rescale it by the extraction fraction, and finally average all of them. In the SM we discuss in details the microwave extraction procedure and the choice of the threshold for the reconstruction method.
From the reconstructed $n_1$ (\aref{fig:fig1}a) we obtain the pressure and density along the long axis \x of the sample.
We independently measure the temperature from the time-of-flight image by fitting the wings of the thermal distribution to a Bose function, taking into account effects due to the non-ballistic expansion from our elongated trap \cite{Szczepkowski2009}.

The pressure of the gas along \x is $p = m \omega_\rho^2 / 2\pi \int n_1\,dy$, where $m$ is the atomic mass. It is obtained integrating the Gibbs--Duhem relation $dp = n d\mu + s dT$ at constant temperature, where $s$ is the entropy density, and assuming the LDA relation $\mu = \mu_0 - V_{ext}$, where $V_{ext}$ is the trapping potential and $\mu_0$ is the value of the chemical potential in the trap center \cite{Cheng2007, Ho2010}.
The {\it in-situ} density can be calculated either from the Gibbs--Duhem relation $n =(\partial p/\partial \mu)_T = - (\partial p/\partial V_{ext})_T$ or from the inverse Abel transform. In Section IV of the SM we provide an explicit comparison between the two methods. Using the Abel transform, we obtain a 2D slice $n(x,y)$ of the density along the imaging plane, that we azimuthally average to obtain a low-noise profile of the density along the \x axis.
Figures \ref{fig:fig1}b and \ref{fig:fig1}c show the pressure and density along \x for a sample of \SI{5.4(5)e6}{atoms} with a temperature  $T = \SI{280 \pm 10}{\nano \kelvin}$, corresponding to a BEC fraction of about \SI{50}{\percent} with a sizable thermal component.
The errorbars in this and in the following figures are due to the uncorrelated combination of statistical and systematic errors on the reconstructed column density, of which we give a detailed description in the SM.

Our configuration is well suited to explore the thermodynamics of the uniform gas in a wide range of densities, that we map to the reduced temperature $T/T_c$, where $T_c=\left(2\pi \hbar^2/m k_B\right)(n/\zeta_{3/2})^{2/3}$ is the local critical temperature. Here $\zeta_{\nu}$ is the Riemann zeta function evaluated at $\nu$. At the trap center, where the density is maximum, we have  $T / T_c \sim 0.2$, while in the thermal tails it rapidly becomes larger than 1.
From the peak density we evaluate the gas parameter $na^3 = 2\times 10^{-6}$. Since $na^3 \ll 1$, the deviations from the predictions of mean field theory are expected to be small, except close to the critical point, as opposed to the ones from IBG which largely fails below the critical point.

\begin{figure}[t]
 \centering
 \includegraphics[]{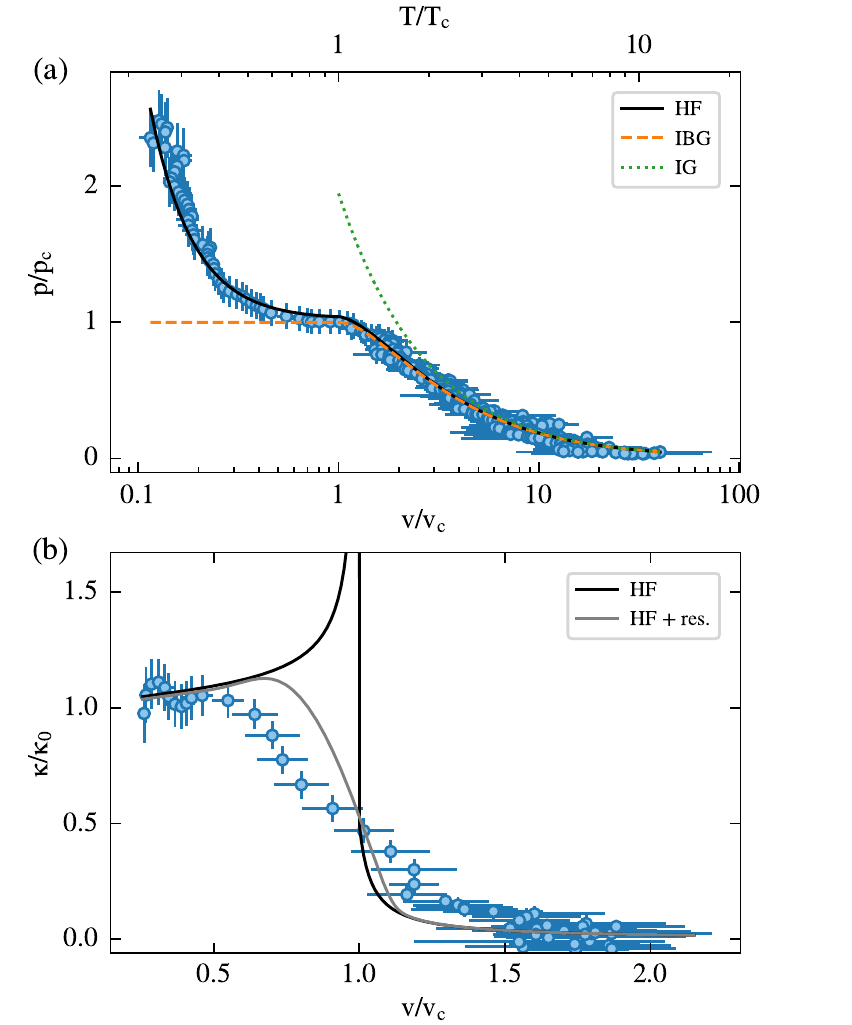}
 \caption{ (a) Measurement of the canonical EoS of a uniform Bose gas, showing the pressure as a function of the specific volume at constant $T = \SI{280}{\nano \kelvin}$. Pressure and density are derived within the LDA from the {\it in-situ} distribution of a harmonically trapped sample. The HF prediction (solid) for a uniform system at the same temperature shows good agreement in the whole range without fitting parameters. Predictions from the IG (dotted) and IBG (dashed) models are shown for comparison. (b) Experimental results for the reduced compressibility versus $v / v_c$ compared to HF theory, which predicts a narrow peak at the critical point. The gray solid line includes the effects of the finite imaging resolution applied to a numerical simulation of the HF density profile in our trap.}
 \label{fig:fig2}
\end{figure}

{\it Canonical EoS $p(v)$ and compressibility.} Figure \ref{fig:fig2}a shows the measurement of the canonical EoS $p(v)$ at constant temperature in reduced variables, rescaled by the relevant critical quantities at the verge of condensation, $p_c = \zeta_{5/2}k_BT/\lambda_T^3$ and $v_c = 1/n_c = \lambda_T^3/\zeta_{3/2}$, where $\lambda_T = (2\pi\hbar^2/m k_B T)^{1/2}$ is the thermal wavelength. The specific volume is related to the local critical temperature by $v/v_c = (T/T_c)^{3/2}$.

The experimental results in \aref{fig:fig2}a are compared to the Hartree--Fock (HF) EoS for uniform matter. The model considers an interacting gas with the following densities for the condensate and thermal fractions, respectively
\begin{equation}
 \label{eq:homog-density-bec}
 \begin{aligned}
    n_0 & = \mu / g - 2 n_T,\\
    n_T & = \frac{1}{\lambda_T^3}g_{3/2}\left(e^{(\mu - 2gn)/k_B T}\right),
 \end{aligned}
\end{equation}
where $g = 4\pi\hbar^2 a / m$, and $g_\nu$ is the polylogarithm function of index $\nu$.
The pressure
\begin{equation}
  \label{eq:homog-pressure}
  p = gn^2 - \frac{1}{2}gn_0^2 + \frac{k_B T}{\lambda_T^3}g_{5/2}\left(e^{(\mu - 2gn)/k_B T}\right)
\end{equation}
can be directly derived from Eqs.~\ref{eq:homog-density-bec}.
The black line in \aref{fig:fig2}a shows the HF EoS evaluated at the experimental value $T = \SI{280}{\nano \kelvin}$, without fitting parameters. We find good agreement between experiment and HF prediction, confirming the validity of the mean field approach for the description of a weakly interacting gas.
For $v/v_c > 1$, the pressure corresponds to that of an ideal (non-interacting) Bose gas (IBG). In the same figure we also show the prediction of the classical ideal gas law $p=k_BT/v$ (IG), which correctly captures the behavior of $p$ only for large $v/v_c$, revealing the importance of quantum effects in the vicinity of the critical point. In the region $v/v_c < 1$, the strong increase in the pressure, that diverges as $(\zeta^2_{3/2}/\zeta_{5/2})(a / \lambda_T)(v_c / v)^2$, shows that the thermodynamics is largely dominated by the effect of the interactions.
The explicit dependence on $a / \lambda_T$ reveals the non-universality of the EoS.

Next, we discuss the isothermal compressibility of the gas defined as $\kappa = (1/n)\,\partial n / \partial p |_T$. Figure \ref{fig:fig2}b shows our measurement of $\kappa$, normalized by the $T=0$ value $\kappa_0 = 1 / gn^2$, as a function of the reduced specific volume.
The experimental results quantitatively agree with the HF prediction (black line) at small $v/v_c$ and show a rapid transition across the critical point. They however strongly deviate from the mean-field prediction in the critical region.

The disagreement can have different origins.
i) The mean field HF theory does not account for the large fluctuation effects characterizing the critical region, which, according to Ginzburg--Landau arguments, corresponds to the range $|\mu - \mu_c| \sim m^3 g^2 k_B^2 T_c^2 / \hbar^6$, with $\mu_c$ the chemical potential evaluated at the transition \cite{Giorgini1996,Arnold2001b}, i.e., to the temperature interval $\Delta T/T\sim an^{1/3}$. 
At the transition $n \sim n_c$ and we have $an_c^{1/3} = (\zeta_{3/2})^{1/3}\,a/\lambda_T \sim \num{5e-3}$. The dependence on $a / \lambda_T$ signals again a violation of universality.
ii) The HF curve is based on the corresponding theory for uniform matter, and does not account for the corrections to the LDA which take place near the transition between the BEC and the normal phase in a trapped system. These lead to a finite thickness of the boundary of the condensate, scaling with the Thomas--Fermi radius $R_x$ as $d = (a_x^4/2R_x)^{1/3}$ \cite{Dalfovo1996a}.
This is a finite-size effect, since $R_x$ depends on the number of atoms in the condensate. We also expect that this result is only weakly affected by the presence of a thermal component. Along the weak axis of our trap $R_x \sim \SI{100}{\micro\meter}$ and we have $d \sim \SI{2}{\micro\meter}$.
iii) Finally, the finite resolution of the imaging system ($\sim \SI{2}{\micro\meter}$) smears the sharp features in the density profile. The gray curve in \aref{fig:fig2}b shows $\kappa$ resulting from a numerical simulation of the column density predicted by HF theory within LDA, and convoluted with our experimental imaging resolution. The remaining differences with respect to the experimental curve are then likely due to the failure of HF theory near the transition and to the violation of the LDA. These are stringent limitations for the measurement of the compressibility in weakly interacting Bose gases as compared to the unitary Fermi gas, where the width of the critical region is much larger as the only energy scale at unitarity is fixed by the Fermi energy, and experiments revealed the occurrence of a peak in $\kappa$.

{\it Chemical potential $\mu(T)$.}
The temperature dependence of $\mu$ in a uniform superfluid system shows a non-monotonic behavior with temperature, which is a peculiar consequence of superfluidity (SM). This feature, already observed in the unitary Fermi superfluid gas \cite{Ku563} has not been so far measured in bosonic ultracold gases.

Within HF theory, $\mu / gn$ increases with temperature in the condensed phase, reaching a peak value of 2 at the transition where $n_T = n$ (see \aref{eq:homog-density-bec}).
The HF approach provides an accurate estimate of the thermodynamics in the temperature range $gn/k_B < T < T_c$.
For $T < gn / k_B$, HF theory ignores the phononic contribution to the thermodynamics, while close to the critical point it neglects the enhanced role of the fluctuations \cite{[{A discussion of the mean-field and beyond-mean-field predictions for the temperature dependence of the chemical potential in a dilute Bose gas, including a comparison with numerical Monte Carlo simulations, is reported in }] ota2020}.

\begin{figure}
 \centering
 \includegraphics[]{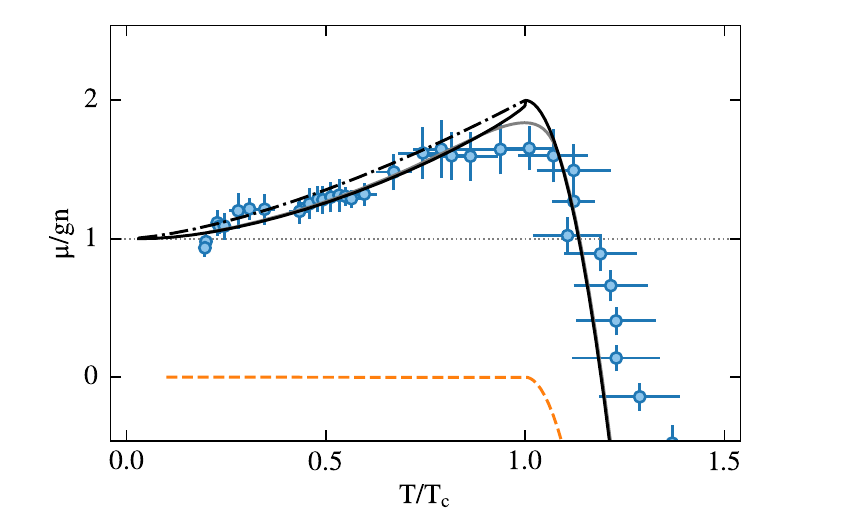}
 \caption{
 Grand canonical EoS for an interacting Bose gas at constant temperature $T = \SI{280}{\nano\kelvin}$. The chemical potential, in units of the local interaction term $gn$, goes from 1 to 2 as the reduced temperature $T/T_c$ goes from 0 to the critical point, and decreases in the thermal region ($T/T_c>1$). 
 Black solid and dot-dashed lines show the HF and $1+(T/T_c)^{3/2}$ predictions, respectively, while the gray solid line includes the effects of the finite imaging resolution applied to a numerical simulation of the HF density profile in our trap. The dashed line is the IBG law.
}
 \label{fig:fig3}
\end{figure}

Within the LDA, the knowledge of $V_{ext}$ is sufficient to measure the chemical potential up to the constant $\mu_0$.
We have determined the value $\mu_0 / k_B = \SI{66.7 \pm 0.2}{\nano \kelvin}$ fitting the density to the HF profile calculated at $T=\SI{280}{\nano\kelvin}$.
The result for $\mu/gn$ as a function of $T/T_c$ is shown in \aref{fig:fig3}, clearly revealing its non-monotonic behavior with a peak around $T = T_c$. In the LDA-based scheme, $T/T_c$ is scanned at fixed $T$ by the spatial variation of $T_c$, which depends on the density. The relevant range across the transition point, $0.2 \leq T/T_c \leq 1.5$, corresponds to the spatial region $\SI{10}{\micro\meter} \leq x \leq \SI{150}{\micro\meter}$.

The discrepancy in the vicinity of the transition is due to the same reasons examined in the analysis of the compressibility. We see that those limitations have a much smaller effect on the determination of $\mu(T)$ compared to \aref{fig:fig2}b, where the dependency on the strong density gradient is more affected by the approximations discussed above.
The dot-dashed line in \aref{fig:fig3} shows the universal curve $\mu / gn = 1 + (T/T_c)^{3/2}$, corresponding to the lowest order approximation for $\mu$ in terms of $g$, obtained by using the IBG result for the thermal fraction $n_T$.
Higher-order corrections to this law stem from the non universality of the Bose gas EoS and affect the exact shape of the curve, but not its non-monotonicity. In Section VII of the SM we present the measurement of the same EoS on an atomic sample at a lower temperature, where the determination of $\mu_0$ does not require the comparison with the HF calculation of the density profile, and find comparable results.
More precise and systematic measurements of the temperature dependence of the chemical potential might provide quantitative estimates of the deviations of $\mu/gn$ from the law $1+(T/T_c)^{3/2}$, caused by the inclusion of beyond mean field effects \cite{ota2020}.

{\it Grand canonical EoS $p(\mu)$}. The above results can be also discussed in the framework of the grand canonical ensemble, where $\mu$ is the independent thermodynamic variable. 
In \cite{Nascimbene2010} it was shown that the dependence of the pressure on the inverse fugacity $\zeta =  e^{-\mu/k_B T}$ reveals a typical cusp behavior at the transition point, with a critical value $\zeta_c \sim 1$, which however could not be measured with sufficient precision to reveal the presence of interaction effects at the transition.
Our analysis shows a critical value of $\zeta_c < 1$, corresponding to a positive shift in the chemical potential in agreement with the results of \aref{fig:fig3}.
The results for the reduced pressure $p/p_c$ as a function of the inverse fugacity are reported and discussed in detail in the SM.

In conclusion, this work contributes to the study of the thermodynamics of a 3D weakly interacting bosonic gas. For the first time we obtain the EoS in the canonical formulation $p(v)$, and highlight the fundamental role of interactions in the finite temperature behavior of a Bose gas.
We provide evidence for the non-monotonic temperature dependence of the chemical potential across the phase transition, a fundamental property which has not been observed before in a weakly interacting superfluid.
Our measurements were possible thanks to the development of an accurate, high-dynamic-range novel imaging method \cite{Mordini2020PTAI}. This approach can be readily applied to other trapped degenerate quantum systems, including the novel phases of interacting quantum mixtures.
Our results lay the groundwork for further investigation of the EoS around the critical region.

\begin{acknowledgments}
We are grateful to M. Ota, S. Giorgini, L. Pitaevskii and C. Salomon for fruitful discussions, and M. Tharrault for contributions at the early stages of this project.
We thank the whole BEC Center and the Q@TN initiative.
We acknowledge funding from the project NAQUAS of QuantERA ERA-NET Cofund in Quantum Technologies (Grant Agreement N. 731473) implemented within the European Union’s Horizon 2020 Programme, and from Provincia Autonoma di Trento.
\end{acknowledgments}

\bibliography{main}

\begin{thebibliography}{29}%
\makeatletter
\providecommand \@ifxundefined [1]{%
 \@ifx{#1\undefined}
}%
\providecommand \@ifnum [1]{%
 \ifnum #1\expandafter \@firstoftwo
 \else \expandafter \@secondoftwo
 \fi
}%
\providecommand \@ifx [1]{%
 \ifx #1\expandafter \@firstoftwo
 \else \expandafter \@secondoftwo
 \fi
}%
\providecommand \natexlab [1]{#1}%
\providecommand \enquote  [1]{``#1''}%
\providecommand \bibnamefont  [1]{#1}%
\providecommand \bibfnamefont [1]{#1}%
\providecommand \citenamefont [1]{#1}%
\providecommand \href@noop [0]{\@secondoftwo}%
\providecommand \href [0]{\begingroup \@sanitize@url \@href}%
\providecommand \@href[1]{\@@startlink{#1}\@@href}%
\providecommand \@@href[1]{\endgroup#1\@@endlink}%
\providecommand \@sanitize@url [0]{\catcode `\\12\catcode `\$12\catcode
  `\&12\catcode `\#12\catcode `\^12\catcode `\_12\catcode `\%12\relax}%
\providecommand \@@startlink[1]{}%
\providecommand \@@endlink[0]{}%
\providecommand \url  [0]{\begingroup\@sanitize@url \@url }%
\providecommand \@url [1]{\endgroup\@href {#1}{\urlprefix }}%
\providecommand \urlprefix  [0]{URL }%
\providecommand \Eprint [0]{\href }%
\providecommand \doibase [0]{http://dx.doi.org/}%
\providecommand \selectlanguage [0]{\@gobble}%
\providecommand \bibinfo  [0]{\@secondoftwo}%
\providecommand \bibfield  [0]{\@secondoftwo}%
\providecommand \translation [1]{[#1]}%
\providecommand \BibitemOpen [0]{}%
\providecommand \bibitemStop [0]{}%
\providecommand \bibitemNoStop [0]{.\EOS\space}%
\providecommand \EOS [0]{\spacefactor3000\relax}%
\providecommand \BibitemShut  [1]{\csname bibitem#1\endcsname}%
\let\auto@bib@innerbib\@empty
\bibitem [{\citenamefont {Smith}\ \emph {et~al.}(2011)\citenamefont {Smith},
  \citenamefont {Campbell}, \citenamefont {Tammuz},\ and\ \citenamefont
  {Hadzibabic}}]{Smith2011}%
  \BibitemOpen
  \bibfield  {author} {\bibinfo {author} {\bibfnamefont {R.~P.}\ \bibnamefont
  {Smith}}, \bibinfo {author} {\bibfnamefont {R.~L.~D.}\ \bibnamefont
  {Campbell}}, \bibinfo {author} {\bibfnamefont {N.}~\bibnamefont {Tammuz}}, \
  and\ \bibinfo {author} {\bibfnamefont {Z.}~\bibnamefont {Hadzibabic}},\
  }\href {\doibase 10.1103/PhysRevLett.106.250403} {\bibfield  {journal}
  {\bibinfo  {journal} {Phys. Rev. Lett.}\ }\textbf {\bibinfo {volume} {106}},\
  \bibinfo {pages} {250403} (\bibinfo {year} {2011})}\BibitemShut {NoStop}%
\bibitem [{\citenamefont {Tammuz}\ \emph {et~al.}(2011)\citenamefont {Tammuz},
  \citenamefont {Smith}, \citenamefont {Campbell}, \citenamefont {Beattie},
  \citenamefont {Moulder}, \citenamefont {Dalibard},\ and\ \citenamefont
  {Hadzibabic}}]{Tammuz2011}%
  \BibitemOpen
  \bibfield  {author} {\bibinfo {author} {\bibfnamefont {N.}~\bibnamefont
  {Tammuz}}, \bibinfo {author} {\bibfnamefont {R.~P.}\ \bibnamefont {Smith}},
  \bibinfo {author} {\bibfnamefont {R.~L.~D.}\ \bibnamefont {Campbell}},
  \bibinfo {author} {\bibfnamefont {S.}~\bibnamefont {Beattie}}, \bibinfo
  {author} {\bibfnamefont {S.}~\bibnamefont {Moulder}}, \bibinfo {author}
  {\bibfnamefont {J.}~\bibnamefont {Dalibard}}, \ and\ \bibinfo {author}
  {\bibfnamefont {Z.}~\bibnamefont {Hadzibabic}},\ }\href {\doibase
  10.1103/PhysRevLett.106.230401} {\bibfield  {journal} {\bibinfo  {journal}
  {Phys. Rev. Lett.}\ }\textbf {\bibinfo {volume} {106}},\ \bibinfo {pages}
  {230401} (\bibinfo {year} {2011})}\BibitemShut {NoStop}%
\bibitem [{\citenamefont {Gerbier}\ \emph {et~al.}(2004)\citenamefont
  {Gerbier}, \citenamefont {Thywissen}, \citenamefont {Richard}, \citenamefont
  {Hugbart}, \citenamefont {Bouyer},\ and\ \citenamefont
  {Aspect}}]{Gerbier2004}%
  \BibitemOpen
  \bibfield  {author} {\bibinfo {author} {\bibfnamefont {F.}~\bibnamefont
  {Gerbier}}, \bibinfo {author} {\bibfnamefont {J.~H.}\ \bibnamefont
  {Thywissen}}, \bibinfo {author} {\bibfnamefont {S.}~\bibnamefont {Richard}},
  \bibinfo {author} {\bibfnamefont {M.}~\bibnamefont {Hugbart}}, \bibinfo
  {author} {\bibfnamefont {P.}~\bibnamefont {Bouyer}}, \ and\ \bibinfo {author}
  {\bibfnamefont {A.}~\bibnamefont {Aspect}},\ }\href {\doibase
  10.1103/PhysRevA.70.013607} {\bibfield  {journal} {\bibinfo  {journal} {Phys.
  Rev. A}\ }\textbf {\bibinfo {volume} {70}},\ \bibinfo {pages} {013607}
  (\bibinfo {year} {2004})}\BibitemShut {NoStop}%
\bibitem [{\citenamefont {Gotlibovych}\ \emph {et~al.}(2014)\citenamefont
  {Gotlibovych}, \citenamefont {Schmidutz}, \citenamefont {Gaunt},
  \citenamefont {Navon}, \citenamefont {Smith},\ and\ \citenamefont
  {Hadzibabic}}]{Gotlibovych2014}%
  \BibitemOpen
  \bibfield  {author} {\bibinfo {author} {\bibfnamefont {I.}~\bibnamefont
  {Gotlibovych}}, \bibinfo {author} {\bibfnamefont {T.~F.}\ \bibnamefont
  {Schmidutz}}, \bibinfo {author} {\bibfnamefont {A.~L.}\ \bibnamefont
  {Gaunt}}, \bibinfo {author} {\bibfnamefont {N.}~\bibnamefont {Navon}},
  \bibinfo {author} {\bibfnamefont {R.~P.}\ \bibnamefont {Smith}}, \ and\
  \bibinfo {author} {\bibfnamefont {Z.}~\bibnamefont {Hadzibabic}},\ }\href
  {\doibase 10.1103/PhysRevA.89.061604} {\bibfield  {journal} {\bibinfo
  {journal} {Phys. Rev. A}\ }\textbf {\bibinfo {volume} {89}},\ \bibinfo
  {pages} {061604(R)} (\bibinfo {year} {2014})}\BibitemShut {NoStop}%
\bibitem [{\citenamefont {Nascimb{\`{e}}ne}\ \emph {et~al.}(2010)\citenamefont
  {Nascimb{\`{e}}ne}, \citenamefont {Navon}, \citenamefont {Chevy},\ and\
  \citenamefont {Salomon}}]{Nascimbene2010}%
  \BibitemOpen
  \bibfield  {author} {\bibinfo {author} {\bibfnamefont {S.}~\bibnamefont
  {Nascimb{\`{e}}ne}}, \bibinfo {author} {\bibfnamefont {N.}~\bibnamefont
  {Navon}}, \bibinfo {author} {\bibfnamefont {F.}~\bibnamefont {Chevy}}, \ and\
  \bibinfo {author} {\bibfnamefont {C.}~\bibnamefont {Salomon}},\ }\href
  {\doibase 10.1088/1367-2630/12/10/103026} {\bibfield  {journal} {\bibinfo
  {journal} {New J. Phys.}\ }\textbf {\bibinfo {volume} {12}},\ \bibinfo
  {pages} {103026} (\bibinfo {year} {2010})}\BibitemShut {NoStop}%
\bibitem [{\citenamefont {Meppelink}\ \emph {et~al.}(2010)\citenamefont
  {Meppelink}, \citenamefont {Rozendaal}, \citenamefont {Koller}, \citenamefont
  {Vogels},\ and\ \citenamefont {van~der Straten}}]{Meppelink2010}%
  \BibitemOpen
  \bibfield  {author} {\bibinfo {author} {\bibfnamefont {R.}~\bibnamefont
  {Meppelink}}, \bibinfo {author} {\bibfnamefont {R.~A.}\ \bibnamefont
  {Rozendaal}}, \bibinfo {author} {\bibfnamefont {S.~B.}\ \bibnamefont
  {Koller}}, \bibinfo {author} {\bibfnamefont {J.~M.}\ \bibnamefont {Vogels}},
  \ and\ \bibinfo {author} {\bibfnamefont {P.}~\bibnamefont {van~der
  Straten}},\ }\href {\doibase 10.1103/PhysRevA.81.053632} {\bibfield
  {journal} {\bibinfo  {journal} {Phys. Rev. A}\ }\textbf {\bibinfo {volume}
  {81}},\ \bibinfo {pages} {053632} (\bibinfo {year} {2010})}\BibitemShut
  {NoStop}%
\bibitem [{\citenamefont {Yefsah}\ \emph {et~al.}(2011)\citenamefont {Yefsah},
  \citenamefont {Desbuquois}, \citenamefont {Chomaz}, \citenamefont
  {G{\"{u}}nter},\ and\ \citenamefont {Dalibard}}]{Yefsah2011}%
  \BibitemOpen
  \bibfield  {author} {\bibinfo {author} {\bibfnamefont {T.}~\bibnamefont
  {Yefsah}}, \bibinfo {author} {\bibfnamefont {R.}~\bibnamefont {Desbuquois}},
  \bibinfo {author} {\bibfnamefont {L.}~\bibnamefont {Chomaz}}, \bibinfo
  {author} {\bibfnamefont {K.~J.}\ \bibnamefont {G{\"{u}}nter}}, \ and\
  \bibinfo {author} {\bibfnamefont {J.}~\bibnamefont {Dalibard}},\ }\href
  {\doibase 10.1103/PhysRevLett.107.130401} {\bibfield  {journal} {\bibinfo
  {journal} {Phys. Rev. Lett.}\ }\textbf {\bibinfo {volume} {107}},\ \bibinfo
  {pages} {130401} (\bibinfo {year} {2011})}\BibitemShut {NoStop}%
\bibitem [{\citenamefont {Desbuquois}\ \emph {et~al.}(2014)\citenamefont
  {Desbuquois}, \citenamefont {Yefsah}, \citenamefont {Chomaz}, \citenamefont
  {Weitenberg}, \citenamefont {Corman}, \citenamefont {Nascimb{\`{e}}ne},\ and\
  \citenamefont {Dalibard}}]{Desbuquois2014}%
  \BibitemOpen
  \bibfield  {author} {\bibinfo {author} {\bibfnamefont {R.}~\bibnamefont
  {Desbuquois}}, \bibinfo {author} {\bibfnamefont {T.}~\bibnamefont {Yefsah}},
  \bibinfo {author} {\bibfnamefont {L.}~\bibnamefont {Chomaz}}, \bibinfo
  {author} {\bibfnamefont {C.}~\bibnamefont {Weitenberg}}, \bibinfo {author}
  {\bibfnamefont {L.}~\bibnamefont {Corman}}, \bibinfo {author} {\bibfnamefont
  {S.}~\bibnamefont {Nascimb{\`{e}}ne}}, \ and\ \bibinfo {author}
  {\bibfnamefont {J.}~\bibnamefont {Dalibard}},\ }\href {\doibase
  10.1103/PhysRevLett.113.020404} {\bibfield  {journal} {\bibinfo  {journal}
  {Phys. Rev. Lett.}\ }\textbf {\bibinfo {volume} {113}},\ \bibinfo {pages}
  {020404} (\bibinfo {year} {2014})}\BibitemShut {NoStop}%
\bibitem [{\citenamefont {Navon}\ \emph {et~al.}(2011)\citenamefont {Navon},
  \citenamefont {Piatecki}, \citenamefont {G\"unter}, \citenamefont {Rem},
  \citenamefont {Nguyen}, \citenamefont {Chevy}, \citenamefont {Krauth},\ and\
  \citenamefont {Salomon}}]{Navon2011}%
  \BibitemOpen
  \bibfield  {author} {\bibinfo {author} {\bibfnamefont {N.}~\bibnamefont
  {Navon}}, \bibinfo {author} {\bibfnamefont {S.}~\bibnamefont {Piatecki}},
  \bibinfo {author} {\bibfnamefont {K.}~\bibnamefont {G\"unter}}, \bibinfo
  {author} {\bibfnamefont {B.}~\bibnamefont {Rem}}, \bibinfo {author}
  {\bibfnamefont {T.~C.}\ \bibnamefont {Nguyen}}, \bibinfo {author}
  {\bibfnamefont {F.}~\bibnamefont {Chevy}}, \bibinfo {author} {\bibfnamefont
  {W.}~\bibnamefont {Krauth}}, \ and\ \bibinfo {author} {\bibfnamefont
  {C.}~\bibnamefont {Salomon}},\ }\href {\doibase
  10.1103/PhysRevLett.107.135301} {\bibfield  {journal} {\bibinfo  {journal}
  {Phys. Rev. Lett.}\ }\textbf {\bibinfo {volume} {107}},\ \bibinfo {pages}
  {135301} (\bibinfo {year} {2011})}\BibitemShut {NoStop}%
\bibitem [{\citenamefont {Cheng}\ and\ \citenamefont {Yip}(2007)}]{Cheng2007}%
  \BibitemOpen
  \bibfield  {author} {\bibinfo {author} {\bibfnamefont {C.-H.}\ \bibnamefont
  {Cheng}}\ and\ \bibinfo {author} {\bibfnamefont {S.-K.}\ \bibnamefont
  {Yip}},\ }\href {\doibase 10.1103/PhysRevB.75.014526} {\bibfield  {journal}
  {\bibinfo  {journal} {Phys. Rev. B}\ }\textbf {\bibinfo {volume} {75}},\
  \bibinfo {pages} {014526} (\bibinfo {year} {2007})}\BibitemShut {NoStop}%
\bibitem [{\citenamefont {Ho}\ and\ \citenamefont {Zhou}(2010)}]{Ho2010}%
  \BibitemOpen
  \bibfield  {author} {\bibinfo {author} {\bibfnamefont {T.-L.}\ \bibnamefont
  {Ho}}\ and\ \bibinfo {author} {\bibfnamefont {Q.}~\bibnamefont {Zhou}},\
  }\href {\doibase 10.1038/nphys1477} {\bibfield  {journal} {\bibinfo
  {journal} {Nat. Phys.}\ }\textbf {\bibinfo {volume} {6}},\ \bibinfo {pages}
  {131} (\bibinfo {year} {2010})}\BibitemShut {NoStop}%
\bibitem [{\citenamefont {Ku}\ \emph {et~al.}(2012)\citenamefont {Ku},
  \citenamefont {Sommer}, \citenamefont {Cheuk},\ and\ \citenamefont
  {Zwierlein}}]{Ku563}%
  \BibitemOpen
  \bibfield  {author} {\bibinfo {author} {\bibfnamefont {M.~J.~H.}\
  \bibnamefont {Ku}}, \bibinfo {author} {\bibfnamefont {A.~T.}\ \bibnamefont
  {Sommer}}, \bibinfo {author} {\bibfnamefont {L.~W.}\ \bibnamefont {Cheuk}}, \
  and\ \bibinfo {author} {\bibfnamefont {M.~W.}\ \bibnamefont {Zwierlein}},\
  }\href {\doibase 10.1126/science.1214987} {\bibfield  {journal} {\bibinfo
  {journal} {Science}\ }\textbf {\bibinfo {volume} {335}},\ \bibinfo {pages}
  {563} (\bibinfo {year} {2012})}\BibitemShut {NoStop}%
\bibitem [{\citenamefont {Shin}\ \emph {et~al.}(2006)\citenamefont {Shin},
  \citenamefont {Zwierlein}, \citenamefont {Schunck}, \citenamefont
  {Schirotzek},\ and\ \citenamefont {Ketterle}}]{Shin2006}%
  \BibitemOpen
  \bibfield  {author} {\bibinfo {author} {\bibfnamefont {Y.}~\bibnamefont
  {Shin}}, \bibinfo {author} {\bibfnamefont {M.~W.}\ \bibnamefont {Zwierlein}},
  \bibinfo {author} {\bibfnamefont {C.~H.}\ \bibnamefont {Schunck}}, \bibinfo
  {author} {\bibfnamefont {A.}~\bibnamefont {Schirotzek}}, \ and\ \bibinfo
  {author} {\bibfnamefont {W.}~\bibnamefont {Ketterle}},\ }\href {\doibase
  10.1103/PhysRevLett.97.030401} {\bibfield  {journal} {\bibinfo  {journal}
  {Phys. Rev. Lett.}\ }\textbf {\bibinfo {volume} {97}},\ \bibinfo {pages}
  {030401} (\bibinfo {year} {2006})}\BibitemShut {NoStop}%
\bibitem [{\citenamefont {Papoular}\ \emph {et~al.}(2012)\citenamefont
  {Papoular}, \citenamefont {Ferrari}, \citenamefont {Pitaevskii},\ and\
  \citenamefont {Stringari}}]{Papoular2012}%
  \BibitemOpen
  \bibfield  {author} {\bibinfo {author} {\bibfnamefont {D.~J.}\ \bibnamefont
  {Papoular}}, \bibinfo {author} {\bibfnamefont {G.}~\bibnamefont {Ferrari}},
  \bibinfo {author} {\bibfnamefont {L.~P.}\ \bibnamefont {Pitaevskii}}, \ and\
  \bibinfo {author} {\bibfnamefont {S.}~\bibnamefont {Stringari}},\ }\href
  {\doibase 10.1103/PhysRevLett.109.084501} {\bibfield  {journal} {\bibinfo
  {journal} {Phys. Rev. Lett.}\ }\textbf {\bibinfo {volume} {109}},\ \bibinfo
  {pages} {084501} (\bibinfo {year} {2012})}\BibitemShut {NoStop}%
\bibitem [{\citenamefont {Freilich}\ \emph {et~al.}(2010)\citenamefont
  {Freilich}, \citenamefont {Bianchi}, \citenamefont {Kaufman}, \citenamefont
  {Langin},\ and\ \citenamefont {Hall}}]{Freilich2010}%
  \BibitemOpen
  \bibfield  {author} {\bibinfo {author} {\bibfnamefont {D.~V.}\ \bibnamefont
  {Freilich}}, \bibinfo {author} {\bibfnamefont {D.~M.}\ \bibnamefont
  {Bianchi}}, \bibinfo {author} {\bibfnamefont {A.~M.}\ \bibnamefont
  {Kaufman}}, \bibinfo {author} {\bibfnamefont {T.~K.}\ \bibnamefont {Langin}},
  \ and\ \bibinfo {author} {\bibfnamefont {D.~S.}\ \bibnamefont {Hall}},\
  }\href {\doibase 10.1126/science.1191224} {\bibfield  {journal} {\bibinfo
  {journal} {Science (80-. ).}\ }\textbf {\bibinfo {volume} {329}},\ \bibinfo
  {pages} {1182} (\bibinfo {year} {2010})}\BibitemShut {NoStop}%
\bibitem [{\citenamefont {Ramanathan}\ \emph {et~al.}(2012)\citenamefont
  {Ramanathan}, \citenamefont {Muniz}, \citenamefont {Wright}, \citenamefont
  {Anderson}, \citenamefont {Phillips}, \citenamefont {Helmerson},\ and\
  \citenamefont {Campbell}}]{Ramanathan2012}%
  \BibitemOpen
  \bibfield  {author} {\bibinfo {author} {\bibfnamefont {A.}~\bibnamefont
  {Ramanathan}}, \bibinfo {author} {\bibfnamefont {S.~R.}\ \bibnamefont
  {Muniz}}, \bibinfo {author} {\bibfnamefont {K.~C.}\ \bibnamefont {Wright}},
  \bibinfo {author} {\bibfnamefont {R.~P.}\ \bibnamefont {Anderson}}, \bibinfo
  {author} {\bibfnamefont {W.~D.}\ \bibnamefont {Phillips}}, \bibinfo {author}
  {\bibfnamefont {K.}~\bibnamefont {Helmerson}}, \ and\ \bibinfo {author}
  {\bibfnamefont {G.~K.}\ \bibnamefont {Campbell}},\ }\href {\doibase
  10.1063/1.4747163} {\bibfield  {journal} {\bibinfo  {journal} {Rev. Sci.
  Instrum.}\ }\textbf {\bibinfo {volume} {83}},\ \bibinfo {pages} {083119}
  (\bibinfo {year} {2012})}\BibitemShut {NoStop}%
\bibitem [{\citenamefont {Mordini}\ \emph {et~al.}(2020)\citenamefont
  {Mordini}, \citenamefont {Trypogeorgos}, \citenamefont {Wolswijk},
  \citenamefont {Lamporesi},\ and\ \citenamefont {Ferrari}}]{Mordini2020PTAI}%
  \BibitemOpen
  \bibfield  {author} {\bibinfo {author} {\bibfnamefont {C.}~\bibnamefont
  {Mordini}}, \bibinfo {author} {\bibfnamefont {D.}~\bibnamefont
  {Trypogeorgos}}, \bibinfo {author} {\bibfnamefont {L.}~\bibnamefont
  {Wolswijk}}, \bibinfo {author} {\bibfnamefont {G.}~\bibnamefont {Lamporesi}},
  \ and\ \bibinfo {author} {\bibfnamefont {G.}~\bibnamefont {Ferrari}},\
  }\href@noop {} {\enquote {\bibinfo {title} {Single-shot reconstruction of the
  density profile of a dense atomic gas},}\ } (\bibinfo {year} {2020}),\
  \Eprint {http://arxiv.org/abs/2005.05674} {arXiv:2005.05674
  [cond-mat.quant-gas]} \BibitemShut {NoStop}%
\bibitem [{\citenamefont {Horikoshi}\ \emph {et~al.}(2017)\citenamefont
  {Horikoshi}, \citenamefont {Ito}, \citenamefont {Ikemachi}, \citenamefont
  {Aratake}, \citenamefont {Kuwata-Gonokami},\ and\ \citenamefont
  {Koashi}}]{Horikoshi2017}%
  \BibitemOpen
  \bibfield  {author} {\bibinfo {author} {\bibfnamefont {M.}~\bibnamefont
  {Horikoshi}}, \bibinfo {author} {\bibfnamefont {A.}~\bibnamefont {Ito}},
  \bibinfo {author} {\bibfnamefont {T.}~\bibnamefont {Ikemachi}}, \bibinfo
  {author} {\bibfnamefont {Y.}~\bibnamefont {Aratake}}, \bibinfo {author}
  {\bibfnamefont {M.}~\bibnamefont {Kuwata-Gonokami}}, \ and\ \bibinfo {author}
  {\bibfnamefont {M.}~\bibnamefont {Koashi}},\ }\href {\doibase
  10.7566/JPSJ.86.104301} {\bibfield  {journal} {\bibinfo  {journal} {J. Phys.
  Soc. Japan}\ }\textbf {\bibinfo {volume} {86}},\ \bibinfo {pages} {104301}
  (\bibinfo {year} {2017})}\BibitemShut {NoStop}%
\bibitem [{\citenamefont {Reinaudi}\ \emph {et~al.}(2007)\citenamefont
  {Reinaudi}, \citenamefont {Lahaye}, \citenamefont {Wang},\ and\ \citenamefont
  {Gu{\'{e}}ry-Odelin}}]{Reinaudi2007}%
  \BibitemOpen
  \bibfield  {author} {\bibinfo {author} {\bibfnamefont {G.}~\bibnamefont
  {Reinaudi}}, \bibinfo {author} {\bibfnamefont {T.}~\bibnamefont {Lahaye}},
  \bibinfo {author} {\bibfnamefont {Z.}~\bibnamefont {Wang}}, \ and\ \bibinfo
  {author} {\bibfnamefont {D.}~\bibnamefont {Gu{\'{e}}ry-Odelin}},\ }\href
  {\doibase 10.1364/OL.32.003143} {\bibfield  {journal} {\bibinfo  {journal}
  {Opt. Lett.}\ }\textbf {\bibinfo {volume} {32}},\ \bibinfo {pages} {3143}
  (\bibinfo {year} {2007})}\BibitemShut {NoStop}%
\bibitem [{\citenamefont {Szczepkowski}\ \emph {et~al.}(2009)\citenamefont
  {Szczepkowski}, \citenamefont {Gartman}, \citenamefont {Witkowski},
  \citenamefont {Tracewski}, \citenamefont {Zawada},\ and\ \citenamefont
  {Gawlik}}]{Szczepkowski2009}%
  \BibitemOpen
  \bibfield  {author} {\bibinfo {author} {\bibfnamefont {J.}~\bibnamefont
  {Szczepkowski}}, \bibinfo {author} {\bibfnamefont {R.}~\bibnamefont
  {Gartman}}, \bibinfo {author} {\bibfnamefont {M.}~\bibnamefont {Witkowski}},
  \bibinfo {author} {\bibfnamefont {L.}~\bibnamefont {Tracewski}}, \bibinfo
  {author} {\bibfnamefont {M.}~\bibnamefont {Zawada}}, \ and\ \bibinfo {author}
  {\bibfnamefont {W.}~\bibnamefont {Gawlik}},\ }\href {\doibase
  10.1063/1.3125051} {\bibfield  {journal} {\bibinfo  {journal} {Rev. Sci.
  Instrum.}\ }\textbf {\bibinfo {volume} {80}},\ \bibinfo {pages} {053103}
  (\bibinfo {year} {2009})}\BibitemShut {NoStop}%
\bibitem [{\citenamefont {Giorgini}\ \emph {et~al.}(1996)\citenamefont
  {Giorgini}, \citenamefont {Pitaevskii},\ and\ \citenamefont
  {Stringari}}]{Giorgini1996}%
  \BibitemOpen
  \bibfield  {author} {\bibinfo {author} {\bibfnamefont {S.}~\bibnamefont
  {Giorgini}}, \bibinfo {author} {\bibfnamefont {L.~P.}\ \bibnamefont
  {Pitaevskii}}, \ and\ \bibinfo {author} {\bibfnamefont {S.}~\bibnamefont
  {Stringari}},\ }\href {\doibase 10.1103/PhysRevA.54.R4633} {\bibfield
  {journal} {\bibinfo  {journal} {Phys. Rev. A}\ }\textbf {\bibinfo {volume}
  {54}},\ \bibinfo {pages} {R4633} (\bibinfo {year} {1996})}\BibitemShut
  {NoStop}%
\bibitem [{\citenamefont {Arnold}\ and\ \citenamefont
  {Tom\'a\ifmmode~\check{s}\else \v{s}\fi{}ik}(2001)}]{Arnold2001b}%
  \BibitemOpen
  \bibfield  {author} {\bibinfo {author} {\bibfnamefont {P.}~\bibnamefont
  {Arnold}}\ and\ \bibinfo {author} {\bibfnamefont {B.}~\bibnamefont
  {Tom\'a\ifmmode~\check{s}\else \v{s}\fi{}ik}},\ }\href {\doibase
  10.1103/PhysRevA.64.053609} {\bibfield  {journal} {\bibinfo  {journal} {Phys.
  Rev. A}\ }\textbf {\bibinfo {volume} {64}},\ \bibinfo {pages} {053609}
  (\bibinfo {year} {2001})}\BibitemShut {NoStop}%
\bibitem [{\citenamefont {Dalfovo}\ \emph {et~al.}(1996)\citenamefont
  {Dalfovo}, \citenamefont {Pitaevskii},\ and\ \citenamefont
  {Stringari}}]{Dalfovo1996a}%
  \BibitemOpen
  \bibfield  {author} {\bibinfo {author} {\bibfnamefont {F.}~\bibnamefont
  {Dalfovo}}, \bibinfo {author} {\bibfnamefont {L.~P.}\ \bibnamefont
  {Pitaevskii}}, \ and\ \bibinfo {author} {\bibfnamefont {S.}~\bibnamefont
  {Stringari}},\ }\href {\doibase 10.1103/PhysRevA.54.4213} {\bibfield
  {journal} {\bibinfo  {journal} {Phys. Rev. A}\ }\textbf {\bibinfo {volume}
  {54}},\ \bibinfo {pages} {4213} (\bibinfo {year} {1996})}\BibitemShut
  {NoStop}%
\bibitem [{\citenamefont {Ota}\ \emph {et~al.}()\citenamefont {Ota},
  \citenamefont {Giorgini},\ and\ \citenamefont {Stringari}}]{ota2020}%
  \BibitemOpen
  \bibfield  {author} {\bibinfo {author} {\bibfnamefont {M.}~\bibnamefont
  {Ota}}, \bibinfo {author} {\bibfnamefont {S.}~\bibnamefont {Giorgini}}, \
  and\ \bibinfo {author} {\bibfnamefont {S.}~\bibnamefont {Stringari}},\
  }\href@noop {} {}\bibinfo {note} {In preparation}\BibitemShut {NoStop}%
\bibitem [{\citenamefont {Dalibard}\ and\ \citenamefont
  {Cohen-Tannoudji}(1985)}]{Dalibard1985}%
  \BibitemOpen
  \bibfield  {author} {\bibinfo {author} {\bibfnamefont {J.}~\bibnamefont
  {Dalibard}}\ and\ \bibinfo {author} {\bibfnamefont {C.}~\bibnamefont
  {Cohen-Tannoudji}},\ }\href {\doibase 10.1088/0022-3700/18/8/019} {\bibfield
  {journal} {\bibinfo  {journal} {J. Phys. B At. Mol. Phys.}\ }\textbf
  {\bibinfo {volume} {18}},\ \bibinfo {pages} {1661} (\bibinfo {year}
  {1985})}\BibitemShut {NoStop}%
\bibitem [{\citenamefont {Holoborodko}(2008)}]{Holoborodko2008}%
  \BibitemOpen
  \bibfield  {author} {\bibinfo {author} {\bibfnamefont {P.}~\bibnamefont
  {Holoborodko}},\ }\href
  {http://www.holoborodko.com/pavel/numerical-methods/numerical-derivative/smooth-low-noise-differentiators/}
  {\enquote {\bibinfo {title} {Smooth noise robust differentiators},}\ }
  (\bibinfo {year} {2008})\BibitemShut {NoStop}%
\bibitem [{\citenamefont {Hansen}\ and\ \citenamefont
  {Law}(1985)}]{Hansen1985}%
  \BibitemOpen
  \bibfield  {author} {\bibinfo {author} {\bibfnamefont {E.~W.}\ \bibnamefont
  {Hansen}}\ and\ \bibinfo {author} {\bibfnamefont {P.-L.}\ \bibnamefont
  {Law}},\ }\href {\doibase 10.1364/JOSAA.2.000510} {\bibfield  {journal}
  {\bibinfo  {journal} {J. Opt. Soc. Am. A}\ }\textbf {\bibinfo {volume} {2}},\
  \bibinfo {pages} {510} (\bibinfo {year} {1985})}\BibitemShut {NoStop}%
\bibitem [{\citenamefont {Gibson}\ \emph {et~al.}(2019)\citenamefont {Gibson},
  \citenamefont {Hickstein}, \citenamefont {Yurchak}, \citenamefont {Ryazanov},
  \citenamefont {Das},\ and\ \citenamefont {Shih}}]{pyabel083}%
  \BibitemOpen
  \bibfield  {author} {\bibinfo {author} {\bibfnamefont {S.}~\bibnamefont
  {Gibson}}, \bibinfo {author} {\bibfnamefont {D.~D.}\ \bibnamefont
  {Hickstein}}, \bibinfo {author} {\bibfnamefont {R.}~\bibnamefont {Yurchak}},
  \bibinfo {author} {\bibfnamefont {M.}~\bibnamefont {Ryazanov}}, \bibinfo
  {author} {\bibfnamefont {D.}~\bibnamefont {Das}}, \ and\ \bibinfo {author}
  {\bibfnamefont {G.}~\bibnamefont {Shih}},\ }\href {\doibase
  10.5281/zenodo.3370021} {\enquote {\bibinfo {title} {Pyabel/pyabel:
  v0.8.3},}\ } (\bibinfo {year} {2019})\BibitemShut {NoStop}%
\bibitem [{\citenamefont {Pitaevskii}\ and\ \citenamefont
  {Stringari}(2016)}]{pitaevskii_stringari_2016}%
  \BibitemOpen
  \bibfield  {author} {\bibinfo {author} {\bibfnamefont {L.}~\bibnamefont
  {Pitaevskii}}\ and\ \bibinfo {author} {\bibfnamefont {S.}~\bibnamefont
  {Stringari}},\ }\href {\doibase 10.1093/acprof:oso/9780198758884.001.0001}
  {\emph {\bibinfo {title} {{Bose-Einstein Condensation and Superfluidity}}}}\
  (\bibinfo  {publisher} {Oxford University Press},\ \bibinfo {year}
  {2016})\BibitemShut {NoStop}%
\end{thebibliography}%

\newpage
\clearpage
\pagebreak
\begin{widetext}
\begin{center}
\textbf{\large Supplemental Materials}
\end{center}
\end{widetext}

\setcounter{equation}{0}
\setcounter{figure}{0}
\setcounter{table}{0}
\setcounter{footnote}{0}
\setcounter{page}{1}
\makeatletter
\renewcommand{\theequation}{S\arabic{equation}}
\renewcommand{\thefigure}{S\arabic{figure}}
\renewcommand{\bibnumfmt}[1]{[S#1]}
\section{Calibration of absorption imaging}

The probe intensity and pulse duration for absorption imaging are chosen in order to optimize the signal-to-noise ratio (SNR) of the optical density, and at the same time to reduce spurious effects coming from atom-light interactions \cite{Horikoshi2017}. Resonant scattering of probe light accelerates the atoms during imaging, which causes a reduction of the optical signal due to Doppler shift of the resonance and a blur of the density distribution due to Brownian motion in the light field \cite{Dalibard1985}. These effects are mitigated by reducing the number of photons scattered per atom during the imaging process, using short and weak probe pulses. On the other hand a poor illumination of the camera results in low SNR images and in a noisy optical density profile. Our choice of the probe light intensity is then motivated by a trade off between these two conditions.
The imaging conditions set also the maximum measurable optical density, at which the illumination of the camera becomes comparable with the noise. For the choice of parameters reported in the main text, we set this threshold to $OD_{th} = 5$.

Scanning the probe light intensity also allows to calibrate the imaging system. The column density $n_1$ is related to the optical density by
\begin{equation}
    \sigma_0 n_1 = \alpha \ln\frac{s_0}{s_1} + s_0 - s_1,
\label{eq:alpha}
\end{equation}
where $\sigma_0$ is the resonant cross section for the atom-light interaction, $s_0 = I_0 / I_{sat}$ is the intensity of the incident probe light and $s_1$ is the intensity transmitted by the atoms, in units of the saturation intensity $I_{sat} = \SI{6.26}{\milli\watt\per\cm^2}$. The coefficient $\alpha$ measures the effective cross section, relative to its value at resonance, and depends on the magnetic field where the atoms are imaged and on the polarization of the probe light.

\begin{figure}
    \centering
    \includegraphics[width=\columnwidth]{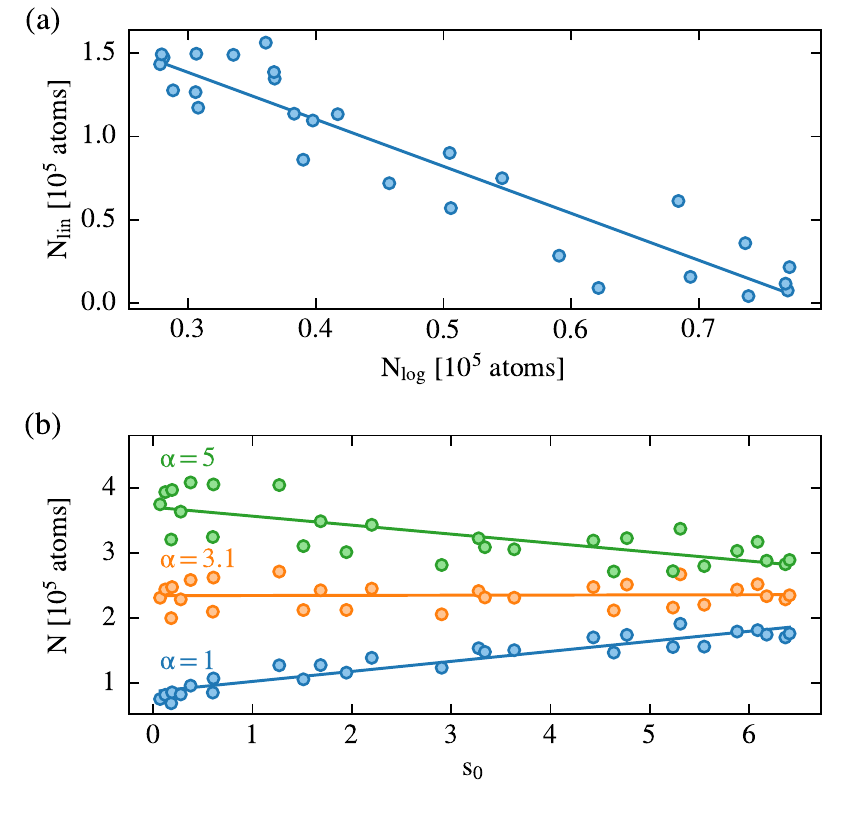}
    \caption{Absorption imaging calibration data. We image in-situ a dilute atomic cloud varying the probe intensity, and the cross section coefficient $\alpha$ is determined so that the measured number of atoms is independent of the illumination conditions.
    (a) We followed the method described in \cite{Horikoshi2017} which finds the value of $\alpha$ that sets the relative weight between the linear and non-linear contributions in the absorption signal, and found $\alpha^{[1]} = \num{3.2 \pm 0.2}$.
    (b) The equivalent method in \cite{Reinaudi2007} minimizes the variation of the measured atom number with the probe intensity, leading to $\alpha^{[2]} = \num{3.10 \pm 0.02}$. The two values are compatible and average to $\alpha = \SI{3.15 \pm 0.12}{}$}
    \label{fig:alpha}
\end{figure}

We measure $s_0$ directly from the camera, by comparing the total pixel count of images of the probe beam with the reading of a calibrated power meter. For the value of $\alpha$, we followed two different (although equivalent) approaches illustrated in \aref{fig:alpha}.
Integrating \aref{eq:alpha} over the region containing the atomic sample and dividing by $\sigma_0$ we reduce it to
\begin{equation}
    N = \alpha N_{log} + N_{lin},
\end{equation}
where $N_{log}$ and $N_{lin}$ are the two contributions to the optical signal coming from the linear and nonlinear absorption regime, respectively. The two quantities depend on $s_0$, but their weighted sum $N$ equals the total number of atoms in the sample and hence must not depend on the details of the probe light.
We measure $N_{log}$ and $N_{lin}$ imaging a dilute atomic sample while scanning $s_0$. On one hand (\aref{fig:alpha}a), we determine $\alpha$ from the slope of a linear fit of $N_{lin}$ versus $N_{log}$ \cite{Horikoshi2017}. On the other hand (\aref{fig:alpha}b), we compute $N$ for several values of $\alpha$ and find the value that minimizes the variation of $N(s_0)$ \cite{Reinaudi2007}, that we effectively extract with a linear fit. The two approaches give comparable values of $\alpha$, that we average to obtain $\alpha = \num{3.15 \pm 0.12}$.

\section{Characterization of the partial transfer}

The novel imaging method we developed combines partial-transfer absorption imaging (PTAI) \cite{Freilich2010,Ramanathan2012} with a high-dynamic-range reconstruction algorithm that allows to image the absolute density of extremely dense trapped atomic samples such as BECs \cite{Mordini2020PTAI}.

\subsection{Calibration}
We compute the fraction of atoms imaged with PTAI from the Rabi frequency of the microwave coupling $\ket{1, -1}$ to $\ket{2, -2}$, which then directly enters in the determination of the absolute density.
As the atomic sample is trapped in a Ioffe--Pritchard magnetic trap, the coupling resonance changes across the sample and in principle it leads to a non-uniform extraction. We modeled this effect from the knowledge of our trap geometry, took it into account in the measurement of the Rabi frequency, and identified the parameter regime where we can neglect the spatial variation of the transfer and avoid distortions in the measured density profile.

\begin{figure}[]
    \centering
    \includegraphics{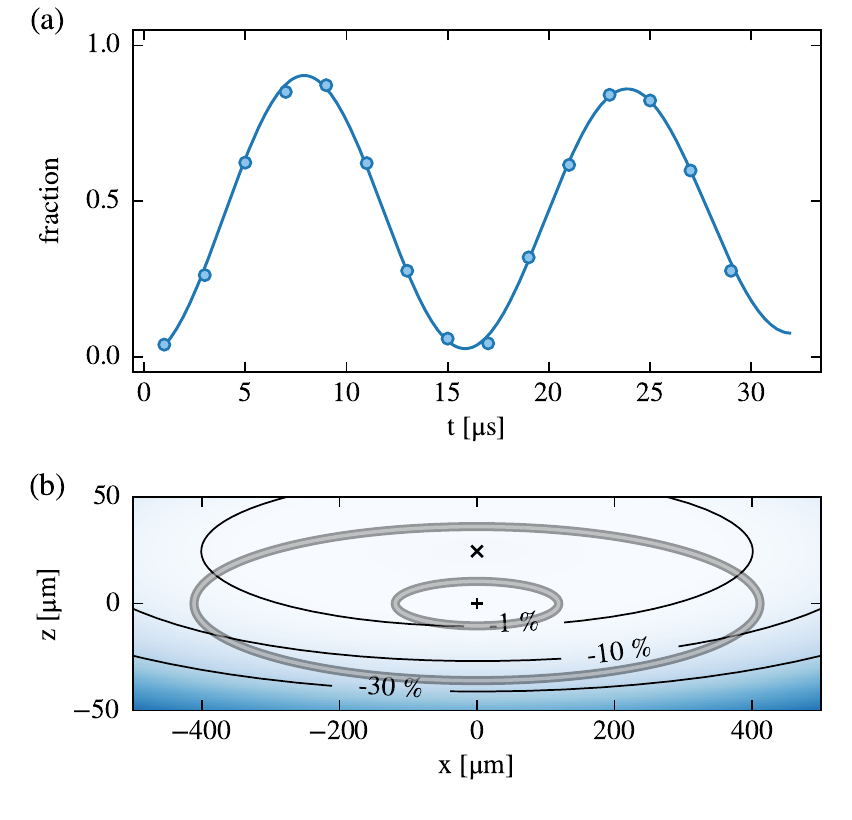}
    \caption{(a) Long-time coherent dynamics of the atomic populations under microwave coupling. The fraction of atoms transferred to $\ket{2, -2}$ is measured in time-of-flight and fitted with the non-uniform transfer model of Eq.\xspace\ref{eq:rabi-approx}. (b) Spatial dependence of $\Delta P / P$ for a microwave pulse of $t = \SI{2.5}{\micro\second}$. The contour lines (black) show the relative deviation with respect to the extraction at resonance.
    The gray ellipses show the boundary of the BEC (inner) and of the thermal component (outer).}
    \label{fig:ptai}
\end{figure}

Due to the combined effect of the trap magnetic field and gravity, the atoms experience the spatially-dependent detuning
\begin{equation}
  \hbar\delta(x, y, z) = \frac{3}{2}m \left( \omega_x^2 x^2 + \omega_\rho^2 (y^2 + z^2 - 2 z z_{sag}) \right),
\label{eq:detuning}
\end{equation}
where $m$ is the atomic mass, $\omega_{x,\rho}$ is the axial (radial) trapping frequency, $z_{sag} = g / \omega_{\rho}^2$ is the gravitational sag, and $g$ is the acceleration of gravity. 
Scanning the microwave frequency we find the value that maximizes the transfer at a given pulsetime, and consider it as the resonant frequency in the center of the atomic cloud, where we set $\delta = 0$.
The local transferred fraction after a pulse of duration $t$ is given by the coherent two-level dynamics
\begin{equation}
    P(t, \delta) = \frac{\Omega^2}{\Omega^2 + \delta^2}\sin^2\left(\frac{t}{2}\sqrt{\Omega^2 + \delta^2}\right)
\label{eq:rabi}
\end{equation}
and it is not uniform as $\delta$ is position-dependent.

To measure the Rabi frequency $\Omega$, we follow the long-time coherent dynamics of the system driven by the microwave field. We apply the microwave for a time $t$, separate the two populations using magnetic field gradients, and image the two atomic clouds. In this way we measure $\tilde P = 1/N \int P(t, \delta(r)) n(r)\,d^3r$, with $n(r)$ the atomic density and $N$ the total atom number, which effectively is a spatial integral of \aref{eq:rabi}.
Since in the region spanned by the atoms the linear term in \aref{eq:detuning} is dominant, we have $\hbar\delta \sim 3 m g z$. Approximating the density distribution to a Gaussian, we can reduce $\tilde P$ to
\begin{equation}
    \tilde P(t) = \frac{1}{\sqrt{2\pi}\Delta_0}\int P(t, \delta)\ e^{-\delta^2/2\Delta_0^2}\, d\delta,
\label{eq:rabi-approx}
\end{equation}
where $\Delta_0$ is the effective range of detuning spanned by the atomic sample. For a condensate of \num{5e6} atoms in our trap, this corresponds to $\Delta_0 / 2\pi \sim \SI{20}{\kilo\hertz}$.
Figure \ref{fig:ptai}a shows a Rabi flop fitted with \aref{eq:rabi-approx}, from which we extract a Rabi frequency $\Omega / 2\pi = \SI{60.7 \pm 0.2}{\kilo\hertz}$.
This effective model correctly describes the effect of the spatial decoherence induced by the magnetic field gradient, and allows us to retrieve the value of the Rabi frequency from long-time oscillations with reduced contrast.

For our imaging method we are instead interested in the short-time dynamics, as we require short microwave pulses to image the high density regions of the sample.
The condition of short pulses is defined by $\sqrt{\Omega^2 + \delta^2} t \ll 1$, so that $P(t, \delta) \simeq (\Omega t /2)^2$ which is effectively independent of $\delta$. To quantify the error introduced by this approximation we compute the relative spatial dependence of the partial transfer
$\Delta P / P = (P(t, \delta) - P(t, 0)) / P(t, 0)$ 
using Eqs. \ref{eq:detuning} and \ref{eq:rabi}. Figure \ref{fig:ptai}b shows the profile of the microwave extraction for a pulse time $t = \SI{2.5}{\micro\second}$, nominally leading to an extraction of \SI{20}{\percent} in the center of the cloud, which is the highest we required for the HDR reconstruction. The ellipses show the region occupied by the atoms, marking the boundary of the BEC (inner) and of the thermal component at $2.5\, \sigma$ (outer).
The relative variation in the extraction profile is less than \SI{1}{\percent} in the region occupied by the condensate, and $\leq \SI{4}{\percent}$ along the \x axis for the whole atomic distribution. Although the deviation becomes significant in the lower side of the cloud, the atomic density in that region is lower than the one in the condensate by a factor of $10^{-2}$, hence its contribution to the optical density is reduced by a similar amount. 
A numerical simulation of the extraction, imaging and reconstruction process shows that the systematic error in the OD introduced by the non-uniform magnetic field is $< \SI{1}{\percent}$ in the region of the \x axis close to the edge of the condensate.

\subsection{Density effects}

Atomic interactions give rise to nonlinear effects in the Rabi dynamics, introducing a systematic error in the calculation of the transferred fraction.
The chemical potential difference of the two coupled populations,
\begin{equation}
    \Delta \mu = n_1  (g_{11} - g_{12}) + n_2  (g_{12} - g_{22}),
\end{equation}
acts as an effective detuning with respect to the bare atomic resonance. Here $g_{ij}=4 \pi \hbar^2 a_{ij} / m$, where the indices $i,j = 1,2$ label the atoms in $\ket{1,-1}$ and $\ket{2,-2}$, with densities $n_1$ and $n_2$, respectively.

To evaluate the magnitude of such nonlinear effects, we set $n_2 = 0$ since we are interested in small transfers. The microwave field is set on resonance with the center of the atomic cloud, so the low density regions are out-of-resonance with respect to the center by the mean-field shift evaluated at the peak density $n_1$. From spectroscopic measurements of the $\ket{1,-1} \to \ket{2,-2}$ microwave transition, we measured such shift to be $\Delta \mu / h \lesssim \SI{5}{\kilo\hertz}$,
from where we get
\begin{equation}
    \frac{\Delta P}{P} \simeq \frac{1}{12}\left(\frac{\Delta \mu}{\hbar}\right)^2 t^2 \lesssim \num{5e-4}
\end{equation}
by a Taylor expansion of \aref{eq:rabi} at small $\delta=\Delta\mu/\hbar$ and for $t = \SI{2.5}{\micro\second}$, which is the pulsetime used to image the low density part of the atomic sample where the effect is stronger.

\begin{figure}[]
    \centering
    \includegraphics{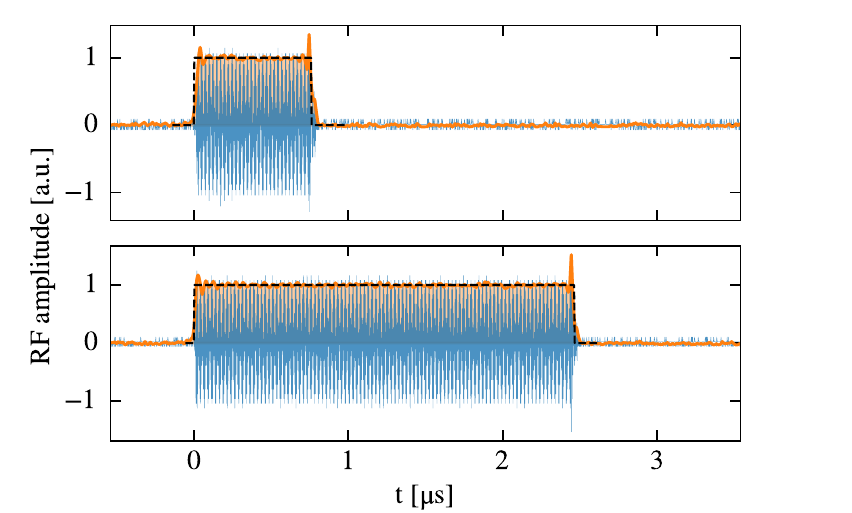}
    \caption{Trace and envelope of microwave pulses of 0.8 (top) and \SI{2.5}{\micro\second} (bottom). The dashed line shows the equivalent (same area) rectangular pulse.}
    \label{fig:pulse-shape}
\end{figure}

\subsection{Pulse shape}

Another source of systematic uncertainties in the transferred fraction is the pulse shape of the microwave field. The effect of a non-rectangular pulse on the population dynamics is to replace $\Omega$ with $(1/t)\int_0^t \Omega(t^\prime)\,dt^\prime$ in \aref{eq:rabi}.
We directly measure the pulse shape of the microwave using a pick-up antenna. Figure \ref{fig:pulse-shape} shows typical traces of pulses with nominal width of 0.8 and \SI{2.5}{\micro\second}. Fitting the pulse envelope we measure a rise/fall time $\tau \sim \SI{20}{\nano\second}$, from which we quantify the deviation from a perfect square pulse as $\tau / t < \SI{3}{\percent}$ for the shortest ones. Nonetheless, we identify the pulse width $t$ as the width of the rectangle with the same area as the actual pulse envelope measured per shot, thus eliminating this systematic source of error.

\section{Error budget}
Table \ref{tab:error-budget} summarizes the contributions to statistic and systematic errors on the reconstructed OD.

The reconstructed image of the atomic cloud is built combining data from many experimental repetitions. The standard deviation across all the images entering the reconstruction measures the total statistical error on the final image, that we measure to be of \SI{4}{\percent} in the region close to the condensate edges.
The number of atoms and the temperature measured in TOF have a relative fluctuation of \SI{3}{\percent}, which quantifies the shot-to-shot stability of our experiment.
We measure the trapping frequencies exciting the dipole mode with a magnetic gradient kick. Their relative uncertainty of \SI{0.5}{\percent} contributes to the errorbars in the pressure profile. We see no higher frequencies in the Fourier spectrum of the oscillation mode, so we assume a purely harmonic potential in the whole region occupied by the atoms.

\begin{table}[h]  
\begin{tabular}{lrr}
source                        & statistic         & systematic     \\
\hline\hline
Reconstruction STD            & \SI{4}{\percent}  & -              \\
Imaging calibration           & -                 & \SI{2.2}{\percent}         \\
Rabi frequency                & -                 & \SI{0.5}{\percent}         \\
Non-uniform magnetic field    & -                 & \SI{1}{\percent}           \\
Density effects               & -                 & $\sim 10^{-4}$   \\
\hline
Total                         & \SI{4}{\percent}  & \SI{2.5}{\percent}
\end{tabular}
\caption{Budget of the statistical and systematic errors on the reconstructed image of the column density.}
\label{tab:error-budget}
\end{table}

\section{Comparison between different methods for the density}

\begin{figure}
    \centering
    \includegraphics[width=\columnwidth]{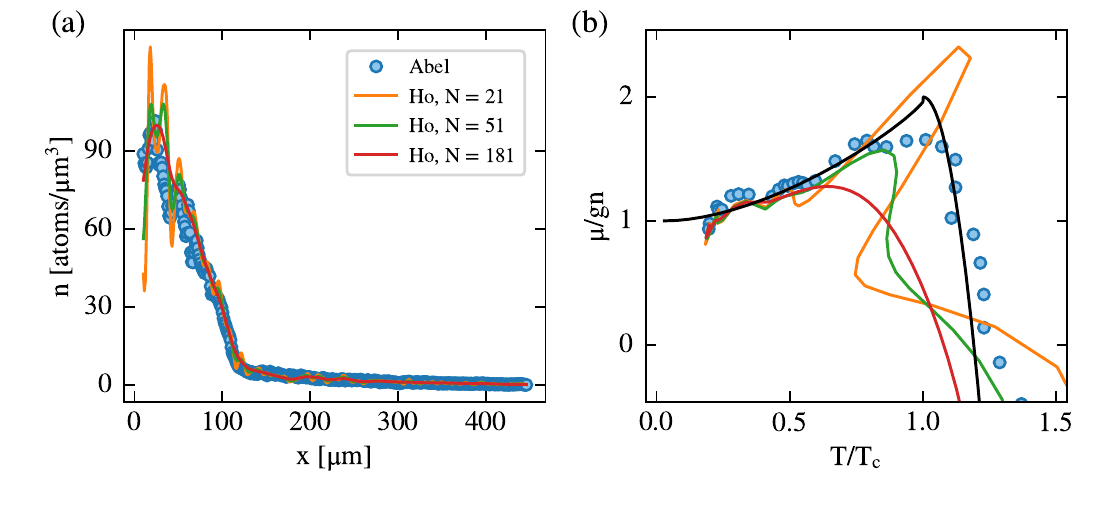}
    \caption{(a) Density profile of a Bose gas calculated with the Abel (dots) or the Gibbs--Duhem method (solid lines) with different filter window lengths $N$. (b) Temperature dependence of the chemical potential obtained from the corresponding density profiles in (a). The black solid line is the HF prediction.}
    \label{fig:Abel-Ho}
\end{figure}

We exploit the cylindrical symmetry of our trapping potential and obtain the 3D density from the inverse Abel transform of the reconstructed column density. We compare this method to the one first proposed in the literature \cite{Ho2010}, based on the Gibbs--Duhem relation and LDA (\emph{GD method} hereafter), which computes $n$ from the derivative 
\begin{equation}
    \frac{dp}{d\mu} = \frac{dp}{dx} \left(\frac{d\mu}{dx}\right)^{-1} = -\frac{1}{m\omega_x^2 x} \frac{dp}{dx}.
\end{equation}
Figure~\ref{fig:Abel-Ho} shows a comparison of the performances of the two different methods. Both are sensitive to the presence of noise, as they involve the calculation of derivatives which amplify the high-frequency components of the signal.
In the GD method, this problem can be mitigated by implementing a suitable smooth numerical differentiator. Generalizations of the central difference scheme are proven to be noise-robust, with stronger smoothing properties depending on the window length $N$ \cite{Holoborodko2008}. In \aref{fig:Abel-Ho}a we show density profiles calculated with different $N$, together with the Abel density profile reported in the main text, while in \aref{fig:Abel-Ho}b we show the corresponding EoS curves for the chemical potential. The smooth derivative requires a large window size to suppress the noise in the pressure profile, effectively averaging the density over distant locations along the \x axis. This, as a side effect, washes out the sharp localized features such as the change in the density slope at the transition point, and has a strong negative impact on the calculation of the EoS.

The use of the Abel transform offers the strong advantage of giving a two-dimensional information: the inverse transform of the column density, which is a 2D projection of the density along the \z axis, is a 2D slice of the density along the $xy$ plane.
This allows to azimuthally average the density slice over iso-density lines, which in the LDA correspond to the elliptical equipotential lines of the harmonic trap. Since beyond-LDA effects modify the density profile depending on the angle and are stronger further away from the axis, we average the data contained in a region within a small angle ($\pm \SI{10}{\degree}$) around the \x axis.
In this way we retrieve a low-noise radial density profile which still preserves its sharp features.
We compute the inverse Abel transform with the Hansen--Law method \cite{Hansen1985, pyabel083}.

\section{Propagation of periodic patterns through the Abel transform}

\begin{figure}[]
 \centering
 \includegraphics{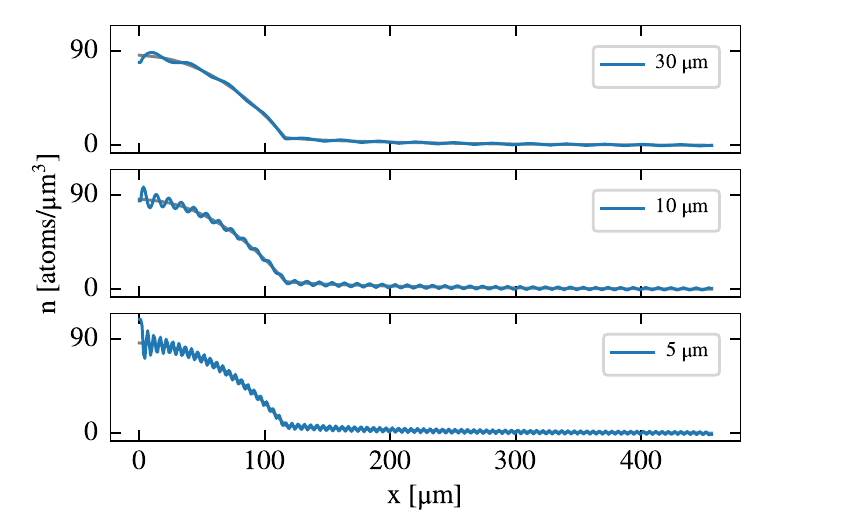}
 \caption{ Simulation of the propagation through Abel inversion of sinusoidal structures on top of a HF column density profile. The shown patterns, from top to bottom, have a period of 30, 10, and \SI{5}{\micro\meter} respectively.}
 \label{fig:fringes}
\end{figure}

The oscillations in the central part of the condensate, that can be seen in Fig. 1 in the main text, are due to residual fringes in the column density produced by the imaging optical system whose effect is amplified in the center of the cloud by the process of Abel inversion. They have not been included in the budget of systematic errors of the reconstructed OD, since they could not be separated from the atomic signal.
To quantify their contribution to the density, we test how such periodic structures are propagated by the Abel inversion when they are added on top of a column integrated Hartree--Fock density profile.
We identify the characteristic amplitude and wavelength of such structures from a Fourier transform of the residuals of a HF fit to the experimental column density, and in \aref{fig:fringes} we show how some of these structures are transformed by the Abel method.
We see that the resulting oscillations that appear on top of the density profile have variable amplitude depending on their periodicity, and are bigger close to the center of the condensate as we consistently observed also in the experimental profile.
We also note that the azimuthal average we perform after the Abel inversion contributes to removing these fringes from the tail of the distribution, but is less effective in the center, where the number of points that are effectively averaged is lower.

\section{Non-monotonous $T$-dependent behavior of the chemical potential}
The non-monotonous behavior of the chemical potential as a function of temperature is a natural consequence of superfluidity \cite{Papoular2012}. In fact, at high temperature in the classical regime, the chemical potential of the gas behaves as
\begin{equation}
\mu(T) \to k_BT \ln{(n \lambda_T^3)}
\end{equation}
and becomes more and more negative as $T\to \infty$. At low temperature its behavior, in  a superfluid, is instead fixed by the thermal excitation of phonons and is given by
\begin{equation}
\mu(T) \to \mu_0 + \frac{\pi^2}{30}\frac{k_BT^4}{\hbar^3c^4}\frac{dc}{dn}
\label{eq:phonons}
\end{equation}
where $\mu_0$ is the value of the chemical potential at $T=0$, while $c$ is the sound velocity at $T=0$. Since at $T=0$ the sound velocity increases with $n$, the chemical potential turns out to be an increasing function of $T$, thereby revealing the non-monotonous behavior as a function of $T$. 
This peculiar non-monotonous effect has been already pointed out experimentally  in the case of a superfluid Fermi gas at unitarity \cite{Ku563} and is also confirmed by the thermodynamic behavior of liquid \he \cite{Papoular2012}. Actually the Hartree--Fock theory developed in the main text for a weakly interacting Bose gas does not account for the phononic behavior of \aref{eq:phonons}, which is expected to hold in the low temperature regime $k_BT\ll \mu_0$,  but is nevertheless consistent with the temperature increase of $\mu(T)$ in the Bose--Einstein condensed region.

\section{Low-temperature measurement}

\begin{figure}[]
 \centering
 \includegraphics{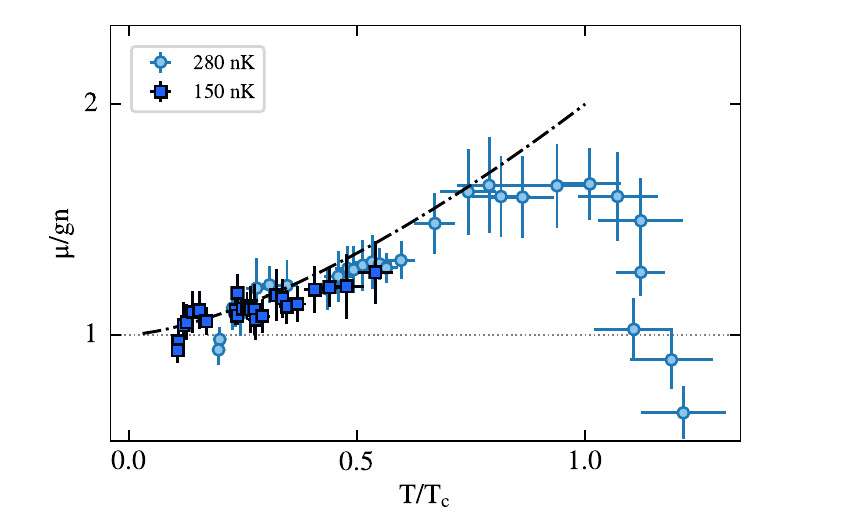}
 \caption{Temperature dependence of the chemical potential, in units of the local interaction term $gn$, for a cold sample at $T = \SI{150}{\nano\kelvin}$ (dark blue squares) and for a $T = \SI{280}{\nano\kelvin}$ one (light blue circles). The dot-dashed line is the universal law $1 + (T / T_c)^{3/2}$.}
 \label{fig:lowT}
\end{figure}

In low-temperature samples the local thermal fraction in the center of the trap is strongly suppressed, therefore we are allowed to neglect its contribution to the peak chemical potential and assume the zero-temperature limit $\mu_0 = gn(0)$. A colder experimental sample then allows us to explore a more restricted region of the EoS, because of the lower density in the thermal component, but without making explicit use of the HF model to retrieve the value of $\mu_0$.

Figure~\ref{fig:lowT} shows our result for $\mu / gn$ versus the local reduced temperature $T/T_c$  for a gas at \SI{150 \pm 5}{\nano \kelvin} and with $\sim \SI{5e6}{atoms}$ (dark blue squares), that we compare to the experimental profile obtained from the higher-temperature sample reported in the main text (light blue circles), and to the universal curve $1 + (T / T_c)^{3/2}$ which neglects beyond mean field effects in the superfluid phase.
A fit to a Thomas--Fermi (TF) profile in the central region of the density ($x < \SI{60}{\micro\meter}$, corresponding to $T/T_c < 0.15$) leads to $\mu_0 = \SI{69.0 \pm 0.4}{\nano \kelvin}$. \textit{A posteriori}, using HF theory we check that the local thermal fraction is $\leq 0.02$ in the whole fitting region, which bounds the systematic error on $\mu_0$. Although the procedure to determine $\mu_0$ neglects the presence of thermal atoms in the center of the condensate, as it assumes a TF model valid at $T = 0$, we observe that the overall behavior of $\mu/gn$ is monotonically increasing. This is in agreement with the HF prediction and with what observed in the sample at higher temperature. It is a clear sign of exchange effects characterizing the thermal contribution to the energy of the system \cite{pitaevskii_stringari_2016}, that we capture even using a zero-temperature model for $\mu_0$.

In this comparison we stress the fact that the two curves, coming from samples at different temperatures, correspond to different values of the parameter $a / \lambda_T$, respectively \num{3e-3} and \num{4e-3} for $T = \SI{150}{\nano\kelvin}$ and \SI{280}{\nano\kelvin}.
This non-universal difference is small, given that both parameters are small and close to each other, and it is not discernible within our measurements. Nonetheless, it does not affect the main point of our observation, the monotonic increase in $\mu$ proper of the finite-temperature behavior in the superfluid phase.

\section{Pressure vs inverse fugacity}

\begin{figure}[b]
 \centering
 \includegraphics{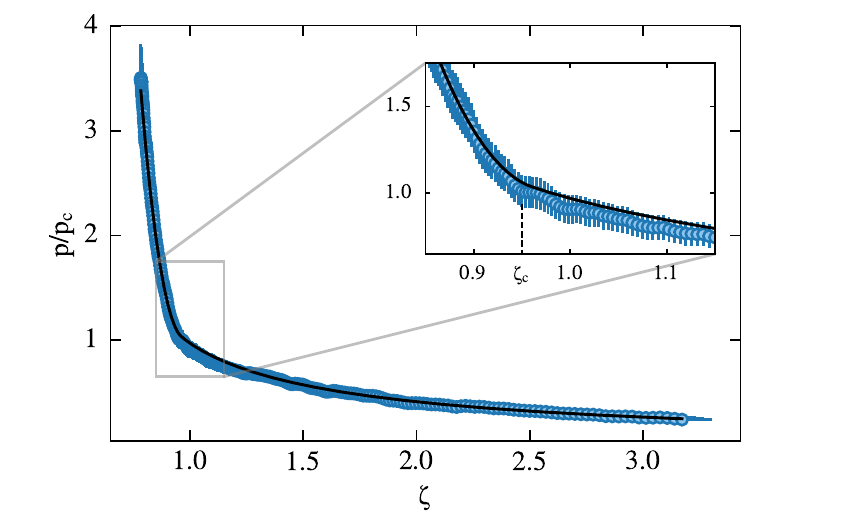}
 \caption{ Grand canonical EoS showing the behavior of the reduced pressure $p/p_c$ as a function of the inverse fugacity. At the transition point, detailed in the inset, the slope changes significantly at a value $\zeta = \zeta_c < 1$ highlighting the mean-field shift of the chemical potential.}
 \label{fig:fugacity}
\end{figure}

Figure~\ref{fig:fugacity} shows the behavior of the reduced pressure $p / p_c$  in terms of the inverse fugacity $\zeta =  e^{-\mu/k_B T}$, along with the HF prediction for a temperature of \SI{280}{\nano\kelvin}.

In the non-degenerate region the pressure does not explicitly depend on $T$ but only on the fugacity, and approaches the IBG prediction $g_{5/2}(\zeta) / \zeta_{5/2}$, defined only for $\zeta \ge 1$. At the transition point, the slope of the pressure profile suddenly increases, signaling the onset of condensation. We observe that the transition happens at $\zeta_c = \SI{0.95 \pm 0.01}{} < 1$, corresponding to a positive chemical potential of $\mu_c = k_B \times \SI{13 \pm 3}{\nano \kelvin}$, which is compatible with the mean-field shift $2 g n_c \simeq k_B \times \SI{12}{\nano \kelvin}$ calculated from the critical density at the given temperature. The effect of the interactions is evident in the deeply degenerate regime $\zeta \to 0$, where the pressure diverges as $(\lambda_T / a) \ln^2 \zeta$. The result of \aref{fig:fugacity} confirms the grand canonical behavior first explored experimentally in \cite{Nascimbene2010}, to which it adds a new observation solidifying the evidence of interaction effects at the transition point.

\end{document}